\documentclass{article}[11pt]
\usepackage{a4wide}
\usepackage{stmaryrd,latexsym,bm,bbm}
\usepackage{epsfig,graphics}
\usepackage{amsmath,amssymb,amsfonts}
\usepackage{color}
\usepackage{epstopdf}
\usepackage{slashed}

\usepackage[font=footnotesize, labelfont=bf]{caption,subcaption}
\usepackage[square,comma,sort&compress,numbers]{natbib}
\usepackage[
colorlinks=true,
linkcolor=black,
breaklinks=true,
urlcolor=blue,
citecolor=green]{hyperref}
\usepackage[utf8]{inputenc}

\usepackage{booktabs}
\usepackage{dcolumn}
\usepackage{makecell}
\usepackage{multirow}

\definecolor{darkblue}{rgb}{0.,0.,0.7}
\definecolor{light-blue}{rgb}{0.8,0.85,1}
\definecolor{green}{rgb}{0,0.6,0}
\definecolor{blueviolet}{rgb}{0.541, 0.169, 0.886}
\definecolor{fuchsia}{rgb}{1.0, 0, 1.0}

\newcommand{\eps}{\epsilon}

\newcommand{\abs}[1]{\lvert#1\rvert}
\newcommand{\mev}{\mathrm{MeV}}
\newcommand{\gev}{\mathrm{GeV}}

\newcommand{\beq}{\begin{equation}}
\newcommand{\eeq}{\end{equation}}
\newcommand{\beqa}{\begin{eqnarray}}
\newcommand{\eeqa}{\end{eqnarray}}

\begin{document}

\title{Hadronic molecular assignment for the newly observed $\Omega^*$ state}
\date{\today}
\author{Yong-Hui Lin$^{1,2,}$\footnote{Email address:
      \texttt{linyonghui@itp.ac.cn} }~   and
      Bing-Song Zou$^{1,2,3}$\footnote{Email address:
      \texttt{zoubs@itp.ac.cn} }
        \\[2mm]
      {\it\small$^1$CAS Key Laboratory of Theoretical Physics, Institute
      of Theoretical Physics,}\\
      {\it\small  Chinese Academy of Sciences, Beijing 100190,China}\\
      {\it\small$^2$University of Chinese Academy of Sciences (UCAS), Beijing 100049, China} \\
      {\it\small$^3$Synergetic Innovation Center for Quantum Effects and Applications (SICQEA)},\\
      {\it\small Hunan Normal University, Changsha 410081, China} \\
}

\maketitle

\begin{abstract}
Very recently, a new $\Omega^{*}$ state was reported by the Belle Collaboration, with its mass of $2012.4 \pm 0.7\ \text{(stat)}\pm 0.6\ \text{(syst)}\ \mev$, which locates just below the $K\Xi^*$ threshold and hence hints to be a possible $K\Xi^*$ hadronic molecule. Using the effective Lagrangian approach as the same as our previous works for other possible hadronic molecular states,  we investigate the decay behavior of this new $\Omega^*$ state within the hadronic molecular picture. The results show that the measured decay width can be reproduced well and its dominant decay channel is predicted to be the $K\pi\Xi$ three-body decay. This suggests that the newly observed $\Omega^*$ may be ascribed as the $J^P={3/2}^-$ $K\Xi^*$ hadronic molecular state and can be further checked through its $K\pi\Xi$ decay channel.
\end{abstract}

\medskip
\newpage

\section{Introduction} \label{sec:1}

Various models, such as classical quenched quark models with three constituent quarks~\cite{Chao:1980em,Capstick:2000qj}, unquenched quark models~\cite{An:2013zoa,An:2014lga} and hadronic dynamical models~\cite{Kolomeitsev:2003kt,Sarkar:2004jh,Si-Qi:2016gmh}, gave very different predictions for the $\Omega^*$ spectrum around 2000 MeV. But experimental knowledge on the $\Omega^*$ spectrum is very pooras listed in the review of the Particle Data Group~\cite{Patrignani:2016xqp}, where the lowest $\Omega^*$ state is $\Omega(2250)$ with its mass about 600 MeV above the $\Omega$ ground state. This is much higher than the predictions of all models for the lowest $\Omega^*$ state.

Very recently, a new $\Omega^{*}$ state was observed in the $\Xi^0 K^-$ and $\Xi^- \bar{K^0}$ invariant mass distributions in $\Upsilon$ decay, by the Belle Collaboration~\cite{Yelton:2018mag}. Its measured mass and decay width are $2012.4 \pm 0.7\ \text{(stat)}\pm 0.6\ \text{(syst)}\ \mev$ and $6.4^{+2.5}_{-2.0}\ \text{(stat)}\pm 1.6\ \text{(syst)}\ \mev$, respectively. The mass is quite close to the
previous quark model prediction of 2020 MeV for the P-wave excitation of the $\Omega$ state~\cite{Chao:1980em}. After the observation of the new $\Omega(2012)$ state, the $qqq$ picture is further explored and supported by the studies with the chiral quark model~\cite{Xiao:2018pwe} and the QCD sum rule method~\cite{Aliev:2018syi}, respectively. On the other hand, the mass is just a few MeV below the $\bar K\Xi(1520)$ threshold of 2015 MeV, which suggests a possible $\bar K\Xi(1520)$ hadron molecule nature for it~\cite{Polyakov:2018mow}, although various previous hadronic dynamical approaches~\cite{Kolomeitsev:2003kt,Sarkar:2004jh,Si-Qi:2016gmh} of the $K\Xi(1520)$ interaction gave very different results.

For the hadronic molecular states, there are many theoretical attempts have been done~\cite{Guo:2017jvc,Chen:2016qju}. A typical example is the pentaquark-like states $P_c^+(4380)$ and $P_c^+(4450)$ observed by LHCb collaboration~\cite{Aaij:2015tga} in 2015. The reported masses of $P_c^+(4380)$ and $P_c^+(4450)$ locate just below the thresholds of $\bar{D}\Sigma_c^*$ and $\bar{D^*}\Sigma_c$ with around $5\ \mev$ and $10\ \mev$ gap, respectively. Inspired by the property that their masses are close to relevant thresholds, our previous work~\cite{Lin:2017mtz} shows that the observed properties of these two $P_c$ states can be reproduced well with the spin-parity-${3/2}^-$ $\bar{D}\Sigma_c^*$ and spin-parity-${5/2}^+$ $\bar{D^*}\Sigma_c$ molecular assumption for $P_c^+(4380)$ and $P_c^+(4450)$ respectively. Actually, it is found that the similar molecular states also exist in strange and beauty sectors~\cite{Lin:2018kcc}. If the new $\Omega(2012)$ state is the S-wave $\bar K\Xi(1520)$ bound state, its spin-parity should be $3/2^-$, just like $P_c(4380)$ as $\bar D\Sigma_c^*$ bound state, $N^*(1875)$ as $K\Sigma^*$ bound state. In the present work, in order to check its hadronic molecular mature, we would like to study the strong decay behaviors of the $\Omega(2012)$ state with the same approach as we did for the $P_c(4380)$ and $N^*(1875)$ states.

This paper is organized as follows: In Sec.~\ref{sec:2}, we introduce formalism and some details about the theoretical tools used to calculate the decay modes of exotic hadronic molecular states. In Sec.~\ref{sec:3}, the numerical results and discussion are presented.

\section{Formalism} \label{sec:2}
With the $\Omega(2012)$ state as the $S$-wave $\Xi(1530)K$ hadronic molecule with spin-parity of ${3/2}^-$, its decay pattern of this molecular state is calculated by means of the effective Lagrangian approach as the same as in our previous work~\cite{Lin:2017mtz,Lin:2018kcc}. The important ingredients of the effective Lagrangian approach are briefly summarized as follows.

At first, the $S$-wave coupling of $\Omega(2012)$ to $\Xi(1530)K$ can be estimated model-independently with the Weinberg compositeness criterion. For the pure hadronic molecular case, it gets that~\cite{Weinberg:1965zz,Baru:2003qq}
\beq\label{eq:coupling}
{g^2} = \frac{4\pi}{4 M m_2}  \frac{(m_1+m_2)^{5/2}} {(m_1 m_2)^{1/2}}
\sqrt{32\epsilon} ,
\eeq
where $M$, $m_1$ and $m_2$ denote the masses of $\Omega(2012)$, $K$ and $\Xi(1530)$,
respectively, and $\eps$ is the binding energy which equals $m_1+m_2-M$. Assuming the physical state in question to be a pure $S$-wave hadronic molecule, the relative uncertainty of the above approximation for the coupling constant is $\sqrt{2\mu\eps}\, r$ where $\mu=m_1 m_2/(m_1+m_2)$ is the reduced mass of the bound particles, and $r$ is the range of forces which may be estimated by the inverse of the mass of the particle that can be exchanged. In our case, $r$ may be estimated as $1/m_{\rho}$.

\begin{figure}[htbp]
\begin{center}
\includegraphics[width=0.5\textwidth]{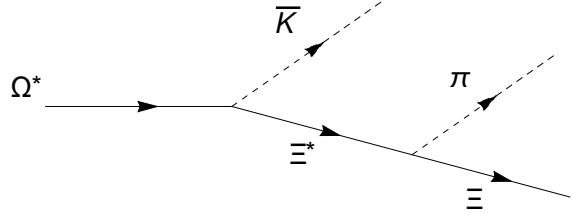}
\caption{The three-body decays of $\Omega(2012)$ in the $\Xi(1530)K$ molecular picture.
\label{Fig:threebody}}
\end{center}
\end{figure}

\begin{figure}[htbp]
\begin{center}
\includegraphics[width=0.5\textwidth]{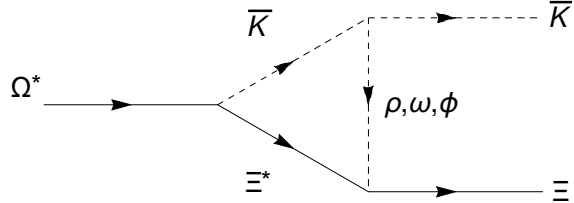}
\caption{The triangle diagram for the two-body decay of the $\Omega(2012)$ in the $\Xi(1530)K$ molecular picture.
\label{Fig:triangle}}
\end{center}
\end{figure}

Note that the decay width of $\Xi(1530)$ listed in PDG is around $9\ \mev$. Compared with the reported width of $\Omega(2012)$, it is apparent that the three-body decay through the decay of $\Xi(1530)$ must be considered during the calculation. However, the four-body decay through the decays of both two constituents is strongly suppressed by the small width of $K$. The dominant three-body decay is given in Fig.~\ref{Fig:threebody}, where the interactions between the final states have been neglected.
To include the contribution of two-body decays, a meson-exchanged triangle diagram convention is taken as the same as our previous work~\cite{Lin:2017mtz,Lin:2018kcc}. For the three-strangeness isospin-zero excited $\Omega^{*}$ molecule, there is only one two-body channel $K\Xi$ need to be considered. The corresponding Feynman diagram is shown in Fig.~\ref{Fig:triangle}. It should be mentioned that the perturbative formalism is used to provide a rough estimation for the total width of $\Omega(2012)$ as we did before, although the nonperturbative approach may be more elegant to give the total widths for a resonance. The partial width is given by
\beq
{\rm d}\Gamma = \frac{F_I}{32 \pi^2} \overline{|{\cal M}|^2}
\frac{|\mathbf{p_1}|}{M^2} {\rm d}\Omega,
\label{eq:widths}
\eeq
where ${\rm d}\Omega = {\rm d}\phi_1 {\rm d}(\cos{\theta_1})$ is the solid angle of particle 1, $M$ is the mass
of the initial $\Omega(2012)$, the factor $F_I$ is from the isospin symmetry, and the
polarization-averaged squared amplitude $\overline{|{\cal M}|^2}$ means $\frac14 \sum_\text{spin} |{\cal M}|^2$.
Note that the types of vertices involved in the amplitudes of the diagrams shown in Fig.~\ref{Fig:threebody} and Fig.~\ref{Fig:triangle} are the same as those that appearing in the processes, spin-parity-$3/2^-$ $K\Sigma^*$ molecule decaying into the $K\pi\Lambda$ and $K\Lambda$ channels. The effective Lagrangians which describe these vertices can be found in our previous papers~\cite{Lin:2017mtz,Lin:2018kcc}. The couplings, $g_{KK\rho}, g_{KK\omega}, g_{KK\phi}, g_{\Xi^*\Xi\rho}, g_{\Xi^*\Xi\omega}$ and $g_{\Xi^*\Xi\omega}$ are taken from the $SU(3)$ relations. The exact values of these couplings used in our calculation are summarized in Table~\ref{table:constants}. And $g_{\Xi^*\Xi\pi}$ is deduced from the experimental decay width of $\Xi(1530)$ decaying into $\Xi\pi$.
\begin{table}[htpb]
\centering
\caption{\label{table:constants}the coupling constants used in the present work. Note that the parameters used in the $\mathrm{SU(3)}$ relations are taken the same values as our previous work. And only absolute values of the couplings are listed with their signs ignored.}
\scalebox{0.85}{
\begin{tabular}{*{11}{c}}
			\toprule\morecmidrules\toprule
			$g_{KK\rho}$ & $g_{KK\omega}$ & $g_{KK\phi}$ & \thead{$g_{\Xi^*\Xi\rho}$ \\ $(\mathrm{GeV}^{-1})$} & \thead{$g_{\Xi^*\Xi\omega}$ \\ $(\mathrm{GeV}^{-1})$} & \thead{$g_{\Xi^*\Xi\phi}$ \\ $(\mathrm{GeV}^{-1})$}	\\
			\Xhline{0.4pt}
			3.02 & 3.02 & 4.27 & 8.44 & 8.44 & 11.94	\\
\bottomrule\morecmidrules\bottomrule 	
\end{tabular}
}
\end{table}

Finally, in order to get rid of the divergence appearing in the loop integration, we take the same convention as our previous work~\cite{Lin:2017mtz,Lin:2018kcc}. The following Gaussian regulator is adopted to suppress short-distance contributions~\cite{Guo:2017jvc,Epelbaum:2008ga,Guo:2013sya,Guo:2013xga,HidalgoDuque:2012pq,Nieves:2011vw,Nieves:2012tt,Valderrama:2012jv},
\beq
f(\bm{p}^2 /\Lambda_0^2) = {\rm{exp}}(-\bm{p}^2 /\Lambda_0^2),
\label{eq:regualtor}
\eeq
where $\bm p$ is the spatial part of the loop momentum and $\Lambda_0$ is an ultraviolet cut-off. During the calculation we vary the $\Lambda_0$ in the range of $0.6-1.4\ \mathrm{GeV}$ to estimate the dependence of our results on the cut-off as we did before. In addition, as described in our previous work a usual form factor chosen as Eq.~\eqref{eq:ff} is also introduced to suppress the off-shell contributions for the exchanged particles.
\beq
f(q^2) = \frac{\Lambda_1^4}{(m^2 - q^2)^2 + \Lambda_1^4},
\label{eq:ff}
\eeq
where $m$ is the mass of the exchanged particle and $q$ is the corresponding momentum. The cut-off $\Lambda_1$ varies
from $0.8\ \mathrm{GeV}$ to $2.0\ \mathrm{GeV}$.

\section{Results and Discussions} \label{sec:3}
With the coupling constants given in Table~\ref{table:constants}, the decay patterns of $\Omega(2012)$ can be calculated numerically. The partial decay widths and the corresponding branch ratios are displayed in Table~\ref{table:decay} with a fixed set of parameters, $\Lambda_0=1.0\ \gev, \Lambda_1=1.2\ \gev$.
\begin{table}[htpb]
\centering
\caption{\label{table:decay}Partial decay widths and branch ratios of $\Omega^{*}(2012)$ with the $S$-wave $\Xi^*K$ molecular scenario. And the cutoffs are fixed as $\Lambda_0=1.0\ \mathrm{GeV}$, $\Lambda_1=1.2\ \mathrm{GeV}$. All of the decay widths are in the unit of $\mathrm{MeV}$, and the short bars denote that the corresponding channel is closed or its contribution is negligible.}
\begin{tabular}{l*{2}{c}}
		\toprule\morecmidrules\toprule
		\multirow{3}*{Mode} & \multicolumn{2}{c}{$J^P={3/2}^-$} \\
		\Xcline{2-3}{0.4pt}
		& \multicolumn{2}{c}{$\Omega^{*}(2012)$\ ($\Xi(1530)K$)}  \\
		\Xcline{2-2}{0.4pt}\Xcline{3-3}{0.4pt}
		& Widths ($\mathrm{MeV}$) & Branch Ratio(\%) \\
		\Xhline{0.8pt}
		$K\Xi$ 	 			& 0.4  	& 14.3 	\\
		$K\pi\Xi$ 	 		& 2.4	& 85.7	\\
		\Xhline{0.8pt}
		Total 				& 2.8	& 100.0	\\
\bottomrule\morecmidrules\bottomrule
\end{tabular}
\end{table}

It should be mentioned that a Breit–Wigner distribution function given by Eq.~\eqref{spectral} is introduced to include the finite width effect of the intermediate state $\Xi^*$ in the three-body decay.
\begin{equation}\label{spectral}
\rho(s)=\frac{N}{\abs{s-m_0^2+i m_0 \Gamma}^2},
\end{equation}
where $m_0$ and $\Gamma$ are the PDG mass and width of $\Xi^*$, respectively. $\sqrt{s}$ is the invariant mass of $\pi\Xi$ final state, varying from $m_0-\Gamma$ to $m_0+\Gamma$. And $N$ is the normalization constant defined as
 \begin{equation}\label{normalize}
 \int_{(m_0-\Gamma)^2}^{(m_0+\Gamma)^2}\rho(s) \mathrm{d}s=1.
 \end{equation}
Note that there is a large and inevitable uncertainty exists in the determination of the coupling constants and the choice of cutoffs $\Lambda_0$ and $\Lambda_1$ in our model. Nevertheless, some qualitative remarks on the decay behaviors of $\Omega^{*}(2012)$ can be obtained from our numerical results. First of all, the small total decay width which is compatible with the announced value is obtained with the $S$-wave $\Xi^*K$ molecular assignment for the reported $\Omega(2012)$. And it is found that the three-body $K\pi\Xi$ decay is the dominant decay channel of $\Omega(2012)$, while the two-body $K\Xi$ channel just contributes 14.3 percent of width at  $\Lambda_0=1.0\ \gev$ and $\Lambda_1=1.2\ \gev$. This is rather different from the prediction of chiral quark model claimed in Ref.~\cite{Xiao:2018pwe}. Future experimental investigation of the three-body decay needs to be performed for disentangling these different assignments of $\Omega(2012)$. Different from the naive expectation of Ref.~\cite{Polyakov:2018mow}, the three-body $K\pi\Xi$ decay width is significantly smaller than the decay width of the free $\Xi(1520)$ state. This is due to the binding energy of the molecule as well as the kinetic energy of $\bar K$ inside the molecule, which reduce the effective mass of the bound $\Xi(1520)$ significantly.  Similar effect was pointed out by Refs.~\cite{Niskanen:2016ntu,Gal:2016bhp} in their studies of $d^*(2380)$ as a $\Delta\Delta$ molecule which gets a decay width smaller than the decay width of a single free $\Delta$ state.

\begin{figure}[htpb]
	\centering
        \includegraphics[width=0.49\textwidth]{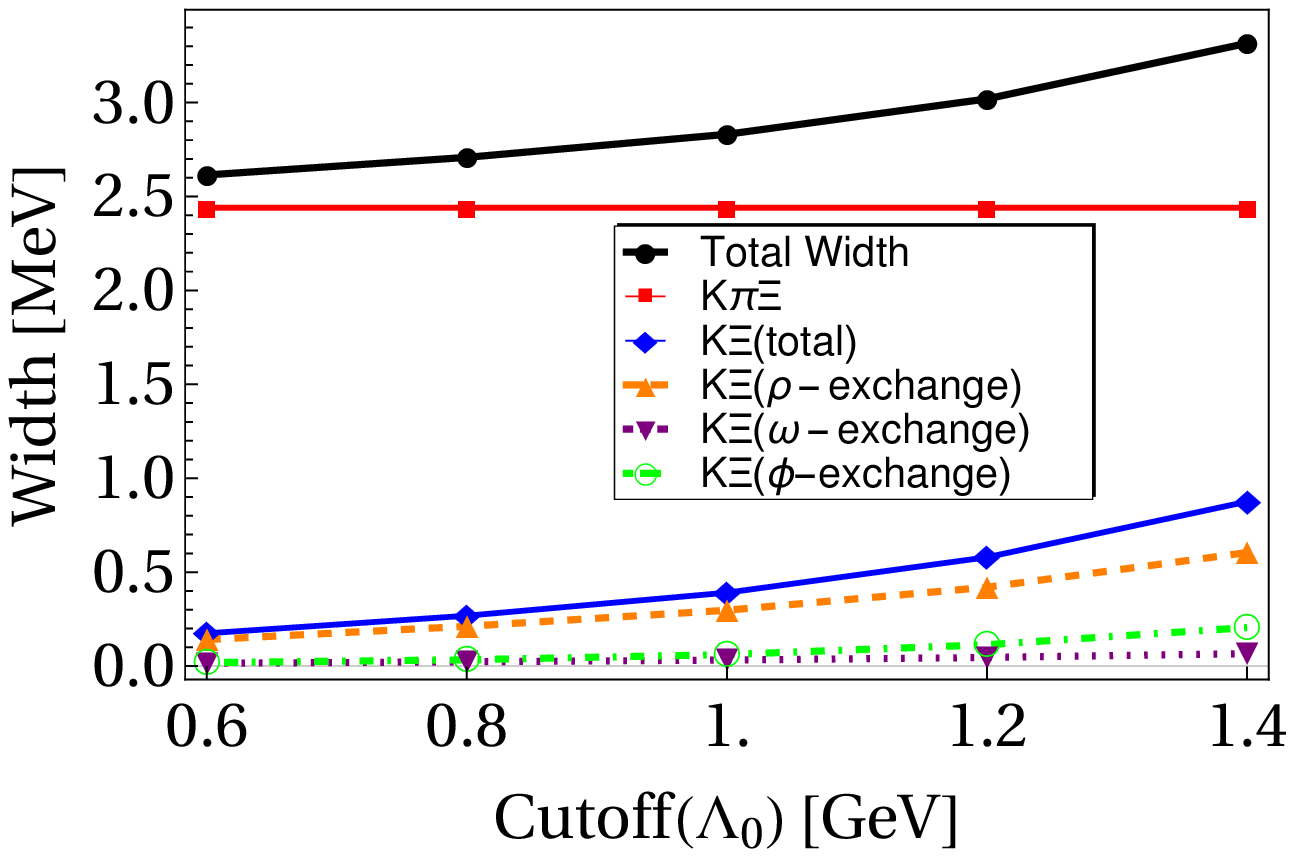}\hfill
        \includegraphics[width=0.49\textwidth]{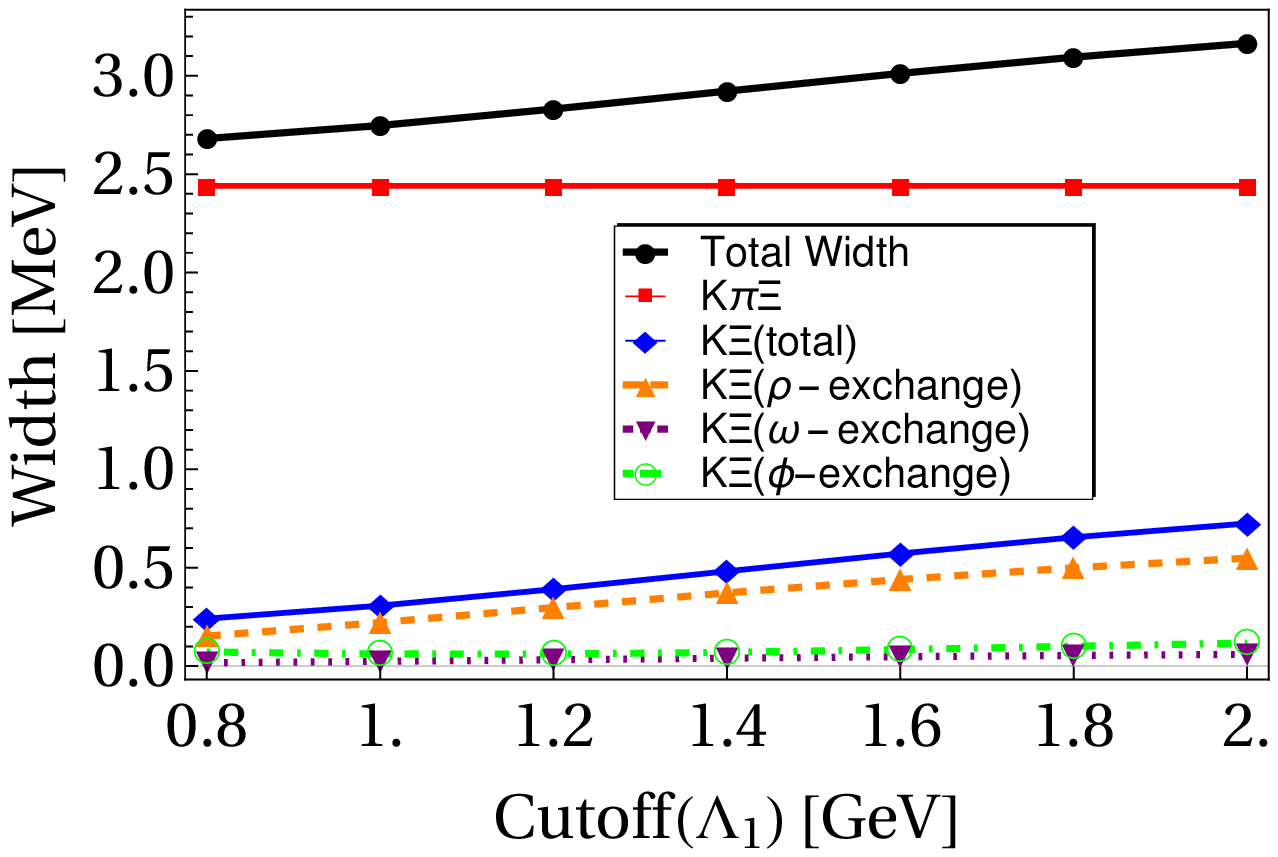}
\caption{\label{figure:partialwidth}
Dependence of the total decay width and partial decay widths of $K\pi\Xi$, $K\Xi$, as well as the partial widths of $\rho$, $\omega$, $\phi$ exchange in the two-body $K\Xi$ decay channel on the cutoffs in the $S$-wave $K\Xi^*$ molecular scenario for $\Omega(2012)$: (left) $\Lambda_0$ changes with $\Lambda_1$ fixed at $1.2\ \gev$; (right) $\Lambda_1$ changes with $\Lambda_0$ fixed at $1.0\ \gev$.}
\end{figure}

\begin{figure}[htpb]
	\centering
	\includegraphics[width=0.95\textwidth]{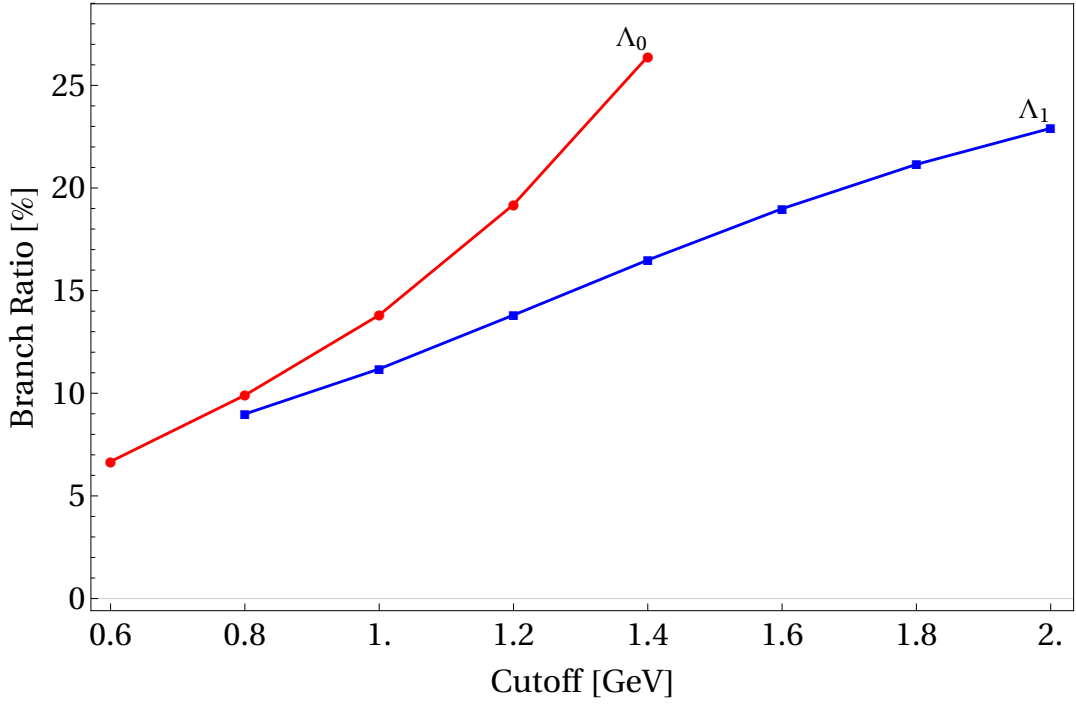}
	\caption{\label{figure:branchratio}
		Dependence of the branch ratio of $K\Xi$ channel on the cutoff $\Lambda_0$ (Red) and $\Lambda_1$ (Blue).}
\end{figure}

\begin{figure}[htpb]
	\centering
        \includegraphics[width=0.95\textwidth]{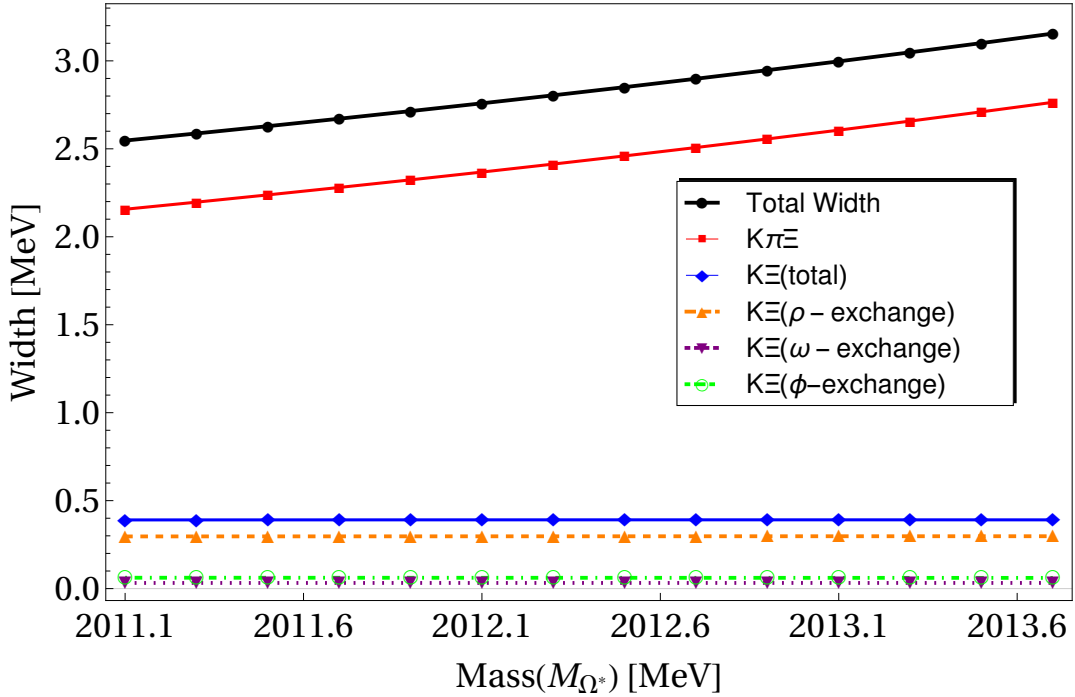}
	\caption{\label{figure:mass}
Dependence of the total decay width and partial decay widths of $K\pi\Xi$, $K\Xi$, as well as the partial widths of $\rho$, $\omega$, $\phi$ exchange in the two-body $K\Xi$ decay channel on the reported mass of $\Omega(2012)$.}
\end{figure}

The cut-off dependence of decay widths is given in Fig.~\ref{figure:partialwidth}. As we can see from the figure, the $\rho$-exchange is the dominant contribution for the partial width of $K\Xi$ two-body channel. And the partial width of three-body $K\pi\Xi$ channel is larger greatly than that of $K\Xi$ channel in the whole ranges of cutoff $\Lambda_0$ and $\Lambda_1$. A measurement of the three-body $K\pi\Xi$ decay branching of the reported $\Omega^*$ candidate will help to test our model and reveal the nature of this new hyperon. The cut-off dependence of the branch ratio of $K\Xi$ channel is shown in Fig.~\ref{figure:branchratio}. Finally, we also analyze the sensitivity of our results to the announced mass of $\Omega(2012)$ as shown in Fig.~\ref{figure:mass}. The curvature shows that the partial width of three-body decay changes slightly within the error bar of reported mass, while the result keeps stable for the $K\Xi$ two-body decay.

In summary, our numerical results indicate that the $S$-wave $\Xi^*K$ molecular scenario for the new $\Omega^*$ candidate can provide a reasonable interpretation for its announced width and the three-body $K\pi\Xi$ decay plays a crucial role on the decay behaviors of $\Omega(2012)$. Searching for this three-body decay of $\Omega(2012)$ can help us to understand its nature.

\bigskip

\section*{Acknowledgments}

We thank Feng-Kun Guo, Jia-Jun Wu and Mao-Jun Yan for helpful discussions. This project is supported by NSFC under Grant No.~11621131001 (CRC110 cofunded by DFG and NSFC) and Grant No.~11747601.


%

\end{document}